\documentclass[preprint,12pt,sort&compress]{elsarticle}

\usepackage{textcomp,booktabs}
\usepackage[usenames,dvipsnames]{color}
\usepackage{colortbl}
\definecolor{mygray}{gray}{.9}
\definecolor{mypink}{rgb}{.99,.91,.95}
\definecolor{mycyan}{cmyk}{.3,0,0,0}
\usepackage{amssymb}
\usepackage{graphicx}
\usepackage{booktabs}
\usepackage{tabularx}
\usepackage{array}
\usepackage{lscape}
\usepackage{longtable}
\usepackage{multirow}
\usepackage{amsmath}
\usepackage{colortbl}
\usepackage[figuresright]{rotating}
\usepackage{epsfig}
\usepackage{epsf}
\usepackage{subfigure}

\newcommand{\PreserveBackslash}[1]{\let\temp=\\#1\let\\=\temp}
\newcolumntype{C}[1]{>{\PreserveBackslash\centering}p{#1}}
\newcolumntype{R}[1]{>{\PreserveBackslash\raggedleft}p{#1}}
\newcolumntype{L}[1]{>{\PreserveBackslash\raggedright}p{#1}}

\journal{} 
\begin{document}

\begin{frontmatter}



\title{Combining Evidences Based on Quantum Mechanical Approach}


\author[address1]{Zichang He}
\author[address1]{Wen Jiang\corref{label1}}
\address[address1]{School of Electronics and Information, Northwestern Polytechnical University, Xi'an, Shaanxi, 710072, China}
\cortext[label1]{Corresponding author at: School of Electronics and Information, Northwestern Polytechnical University, Xi'an, Shaanxi 710072, China. Tel: +8613363912605. E-mail address: jiangwen@nwpu.edu.cn, jiangwenpaper@hotmail.com}

\begin{abstract}
Dempster-Shafer evidence theory is wildly applied in multi-sensor data fusion. However, lots of uncertainty and interference exist in practical situation, especially in the battle field. It is still an open issue to model the reliability of sensor reports. Many methods are proposed based on the relationship among collected data. In this letter, we proposed a quantum mechanical approach to evaluate the reliability of sensor reports, which is based on the properties of a sensor itself. The proposed method is used to modify the combining of evidences.
\end{abstract}


\end{frontmatter}
\section{Introduction}\label{Introduction}
%% Belief function广泛应用于军事，信息融合是open issue
%% 前人的工作都基于数据本身的信息
%% 我们提出一种基于数据源的融合方法
Data fusion has been widely studied in the last decades, especially its military applications. Multi-sensor data fusion (MSDF) technology plays a more and more significant role for the fighting demand. How to fuse the sensor data is still an open issue\cite{7823095,7370554,Jiang2016Sensor,7370262}. %Many basic approaches has been developed, such as Bayesian theory, fuzzy logic, Dempster-Shafer (DS) evidence theory and so on.
Due to the powerful ability of handling uncertain information, DS evidence theory is widely used in MSDF\cite{Basir2007Engine,Dallil2013Sensor,Maherin2015Multistep,Weeraddana2015Dempster}. However, Lots of interference exist in the complex practical situation. The information provided by a sensor report is likely to be disturbed and incorrect. In this case, strong conflict may exist among evidences and lead to a wrong fusion result. Handling conflict is crucial in data fusion\cite{M2016Information,Perez2016Using,Zhao2016125,Zhang201788}. To address it, many approaches have been proposed\cite{Wang2016Combination,Yager2016Soft,Jiang2017mGCR}.

To deal with conflictive information, most previous methods handle evidences based on the relationship among the data collected by sensors\cite{Deng2015Evidence,Liu2006Analyzing,Murphy2000Combining,Guo2006Evaluating}. In this letter, however, an method which bases on the properties of a sensor itself is proposed. To evaluate the reliability of sensor reports, a confidence coefficient curve is determined based on a quantum mechanical approach. Interest in quantum approach to classical fuzzy logic has increased over the last decades\cite{abd2015novel,Bolotin2001Quantum,Dubois2016Eigenlogic,Jaros2015Quantum}.
In classical mechanics, a particle is located in an exact place. If a particle is known to be in M, then it can never in any other places, like in N. In quantum mechanics, however, a particle can never be exactly located due to the well-known Heisenberg's uncertainty relation. Only the probability of finding the particle in a given area like M or N can be determined (shown as Figure \ref{particle}). This interesting property of quantum mechanics is used to describe the reliability degree of a sensor report as it is hard to assert that one sensor report is totally reliable or unreliable. Then we use the curve to calculate the credibility of evidences. The fusion results of the modified evidences show the effectiveness of our method.

\section{Preliminaries}
Dempster-Shafer evidence theory was proposed by Dempster in 1967\cite{Dempster1967Upper} and modified by Shafer in 1978\cite{Shafer1978A}. In evidence theory, the basic set $\Theta$, called the frame of distribution, consists of a set of $N$ mutually exclusive and exhaustive hypotheses, symbolized by $\Theta {\rm{ = }}\left\{ {{X_{\rm{1}}},{X_2}, \ldots ,{X_N}} \right\}$
Let $P\left( \Theta  \right)$ denote the power set composed of ${{\rm{2}}^N}$ elements of $\Theta$.
\[P\left( \Theta  \right){\rm{ = }}\left\{ {\emptyset ,\left\{ {{X_{\rm{1}}}} \right\},\left\{ {{X_2}} \right\}, \ldots ,\left\{ {{X_N}} \right\}, \ldots ,\left\{ {{X_1} \cup {X_2}} \right\},\left\{ {{X_1} \cup {X_3}} \right\}, \ldots ,\Theta } \right\}\]
Basic probability assignment (BPA) is a mapping from $P\left( \Theta  \right)$ to $\left[ {0,1} \right]$, defined by:
\begin{equation}
m: P\left( \Theta  \right) \to \left[ {0,1} \right]
\end{equation}
satisfying the following conditions:
\begin{equation}
\sum\limits_{A \in {2^N}} {m\left( A \right)}  = 1
\end{equation}
\begin{equation}
m\left( \emptyset  \right) = 0
\end{equation}
The mass function $m$ represents a supporting degree to $A$. The elements of $P\left( \Theta  \right)$ that have a non-zero mass are called focal elements. A body of evidence (BOE) is the set of all the focal elements\cite{Jousselme2001A}:
\[\left( {R,m} \right) = \left\{ {\begin{array}{*{20}{c}}
{\left[ {A,m\left( A \right)} \right];}&{A \in }
\end{array}P\left( \Theta  \right){\kern 1pt} {\kern 1pt} {\kern 1pt} {\kern 1pt} {\kern 1pt} and{\kern 1pt} {\kern 1pt} {\kern 1pt} {\kern 1pt} {\kern 1pt} m\left( A \right) > {\rm{0}}} \right\}\]
$R$ is a subset of $P\left( \Theta  \right)$, and each of $A \in P\left( \Theta  \right)$ has a fixed value. The classical Dempster's combining rule of two BOE ${m_1}$ and ${m_2}$ is defined as following:
\begin{equation}\label{combiningrule}
m\left( A \right) = \frac{{\sum\nolimits_{B \cap C = A} {{m_1}\left( B \right){m_2}\left( C \right)} }}{{1 - K}}
\end{equation}
where $K$ is called conflict coefficient:
\begin{equation}
K = \sum\limits_{B \cap C = \emptyset } {{m_1}\left( B \right){m_2}\left( C \right)}
\end{equation}

\section{Quantum mechanical modelling of the sensor reliability in data fusion}
Radar plays an important role in the modern battlefield. Usually, to obtain the overall information, data from several radars need to be fused. 
Aiming to do a more reasonable fusion, we propose an method based on quantum mechanics to determine the confidence coefficient curve of radar sensor reports. We assume that the reliability of sensor reports relates to the distance between object and sensor in some degrees. For each distance $x$, the sensor has an according confidence coefficient whose maximum value is 1. Hence, confidence coefficient curve $\mu \left( x \right)$  is defined as a function to describe this relationship.

The signal of the object is received by $k$ radars. The transmit power of the object is ${P_t}$, the antenna gain of the object is ${G_t}$, the antenna gain of the reconnaissance radar is ${G_r}$, the distance between object and a radar is denoted as $x$. The signal power received by radar is:
\begin{equation}\label{P_r}
{P_r} = \frac{{{P_t}{G_t}{G_r}\sigma {\lambda ^2}}}{{{{\left( {4\pi x} \right)}^2}}}
\end{equation}
where $\lambda$ is the wavelength and $\sigma$ is Radar Cross-Section which is the product of geometric cross-section, reflection coefficient and direction coefficient.

\begin{figure}[!ht]
\centering
\includegraphics[width=3in]{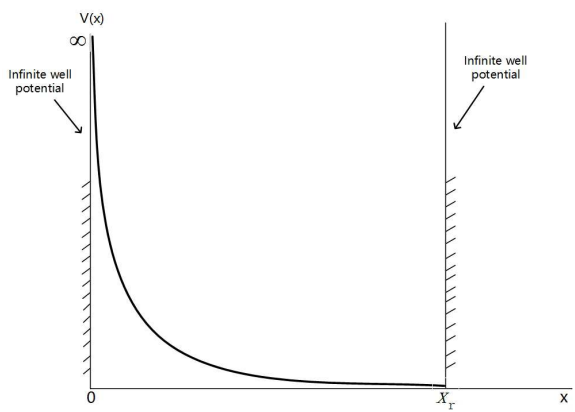}
\caption{The quasi-potential function $V\left( x \right)$}\label{potential}
\end{figure}

If the sensitivity of a radar is ${P_{r\min }}$, the maximal reconnaissance distance ${x_r}$ is calculated as follows. If the object is far beyond this distance, it will not be effectively reconnoitred.
\begin{equation}
{x_r} = {\left[ {\frac{{{P_t}{G_t}{G_r}\sigma {\lambda ^2}}}{{{{\left( {4\pi } \right)}^2}{P_{r\min }}}}} \right]^{{\textstyle{1 \over 2}}}}
\end{equation}
According to the quantum-mechanical rules of quantification, we should write an operator which corresponds to the received signal power:
\begin{equation}
H =  - {c^2}\frac{{{\partial ^2}}}{{\partial {x^2}}} - V\left( x \right)
\end{equation}
where $c$ is a scale factor. $V\left( x \right)$ is a quasi-potential function to model the received power.
\begin{equation}
V\left( x \right)=\left\{ \begin{array}{cc}
{  \frac{\gamma }{{{x^2}}}} & {0 < x \le {x_r}} \\
\infty  & x \le 0, x > {x_r} \\
\end{array} \right.
\end{equation}
where $\gamma  \propto \frac{{{P_t}{G_t}{G_r}\sigma {\lambda ^2}}}{{{{\left( {4\pi } \right)}^2}}}$ corresponds to the parameters in Eq. (\ref{P_r}). The quasi-potential function $V\left( x \right)$ is roughly illustrated as Figure \ref{potential}.

Based on quantum-mechanical rules, a quasi time-independent Schr${\ddot o}$dinger equation can be obtained.
\begin{equation}\label{shcordinger}
H\psi \left( x \right) = L\psi \left( x \right)
\end{equation}
where $L$ relates to the level of the radar sensitivity ${P_{r\min }}$.

The solution of Eq. (\ref{shcordinger}) is a quasi-amplitude distribution $\psi \left( x \right)$.
When the object is within the maximal reconnaissance distance ${x_r}$, we can obtain:
\begin{equation}
\psi \left( x \right) \propto \sqrt x \left[ {{J_\alpha }\left( {\frac{{\sqrt L }}{c}} \right) + {Y_\alpha }\left( {\frac{{\sqrt L }}{c}} \right)} \right]
\end{equation}
where ${{J_\alpha }}$ and ${{Y_\alpha }}$ are the Bessel function of the first kind and the second kind respectively. $\alpha$ is their order:
\begin{equation}
\alpha  = \frac{1}{2}\sqrt {\frac{{{c^2} - 4\gamma }}{{{c^2}}}}
\end{equation}
Then let us consider the other situation, when the object is beyond ${x_r}$, the value of $V\left( x \right)$ is infinite. According to quantum mechanics, it is impossible for a particle to penetrate the well wall if it is within a infinite well potential. Hence, we can conclude that $\psi \left( x \right)=0$ in this case.

\begin{figure}[!ht]
\centering
\includegraphics[width=3in]{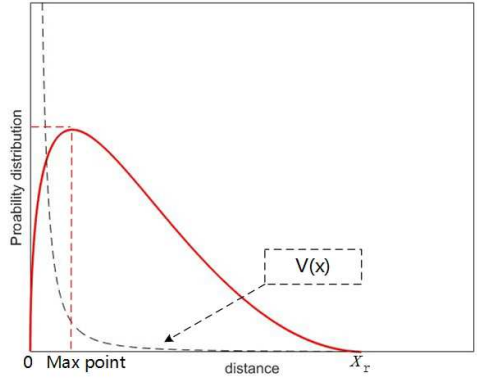}
\caption{The probability distribution $P\left( x \right)$ }\label{px}
\end{figure}

Then we can obtain the probability distribution $P\left( x \right)$, which is illustrated graphically in Figure \ref{px}.
\begin{equation}\label{Pequation}
P\left( x \right) = {\left| {\psi \left( x \right)} \right|^2} \propto x{\left[ {{J_\alpha }\left( {\frac{{\sqrt L }}{c}} \right) + {Y_\alpha }\left( {\frac{{\sqrt L }}{c}} \right)} \right]^2}
\end{equation}

\begin{figure}[!ht]
\centering
\includegraphics[width=3in]{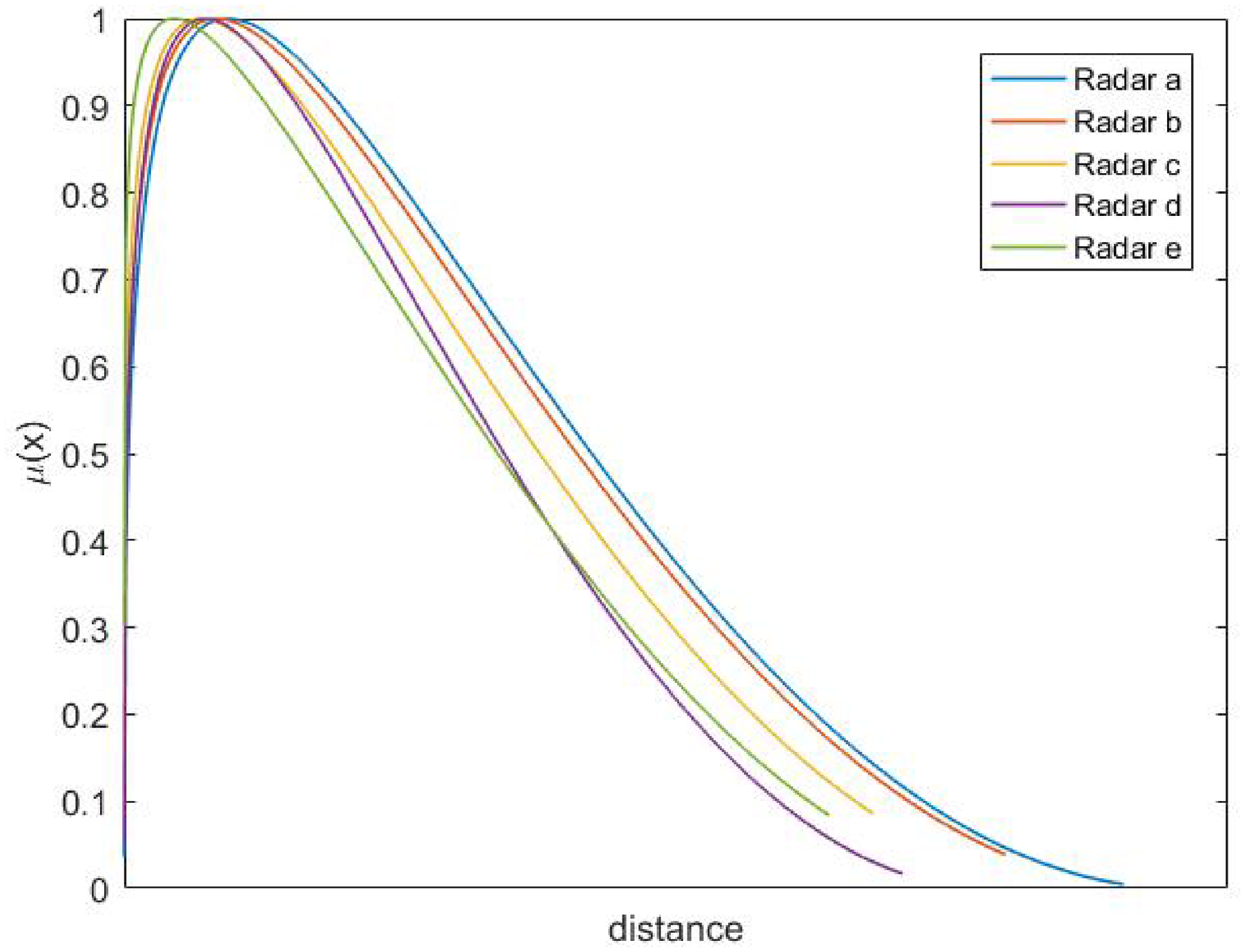}
\caption{The confidence coefficient curves of different radars}\label{multi}
\end{figure}

By amplifying Eq. (\ref{Pequation}), we can obtain the confidence coefficient curve $\mu \left( x \right)$.
\begin{figure}[!ht]
\centering
\includegraphics[width=3in]{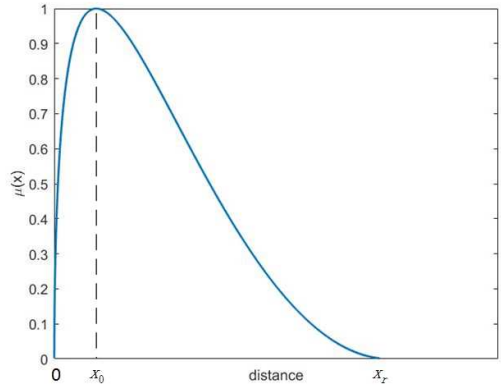}
\caption{The confidence coefficient curve $\mu \left( x \right)$}\label{mux}
\end{figure}
Seen from Figure \ref{mux}, the curve rises rapidly when $x$ is smaller than ${x_0}$ and comes to its maximum when $x$ equals to ${x_0}$. Then it declines slowly until $x$ comes to ${x_r}$, which is reasonable. In practical situation, due to precision and some other intricate issues, a radar do not work well when it is too close to the object. There exists an optimal distance ${x_0}$ for a radar to work. Then the performance of a radar becomes poorer as it is located further. When the distance is further than the maximal reconnaissance distance, the radar can not reconnoitre the object effectively. With the basis of this curve, we can evaluate the reliability of radar reports effectively. For different types of radars, we can obtain their according confidence coefficient curves as Figure \ref{multi}. The parameters of these curves are in Table \ref{curves}.
\begin{table}[!h]
\centering
\caption{Curves of different radars}
\label{curves}
\begin{tabular}{cccc}
\toprule
        & $c$ & $L$  & $r$  \\
\midrule
 Radar a& 10    &0.7  & 14 \\
 Radar b& 10  &0.8  & 12\\
 Radar c& 10  &1.0 & 10 \\
 Radar d& 10    &1.1  & 13 \\
 Radar e& 10  &1.3  & 6 \\
 \bottomrule
\end{tabular}
\end{table}
In the following, the curves are used in combining evidences.
Assume we have $k$ pieces of BOEs: ${m_1},{m_2}, \ldots ,{m_k}$, collected from $k$ radar sensors. By using confidence coefficient curves, each BOE corresponds to one confidence coefficient: ${\mu _1},{\mu _2}, \ldots ,{\mu _k}$. The credibility degree ${Cr{d_i}}$ of BOE ${m_i}$ is defined as:
\begin{equation}
{Cr{d_i}} = \frac{{{\mu _i}}}{{\sum\limits_{i = 1}^k {{\mu _i}} }}
\end{equation}
It is easy to find that $\sum\nolimits_{i = i}^k {{Cr{d_i}}} $. Hence, the credibility degree reveals the relatively importance of the collected evidence.
After determining the credibility of each BOE, we do a modified average for all $k$ pieces of BOEs to obtain a new evidence ${m'}$.
\begin{equation}
m' = \sum\limits_{i = 1}^k {Cr{d_i} \times {m_i}}
\end{equation}
Then we can combine ${m'}$ with itself for $k-1$ times by using classical combining rule (Eq. (\ref{combiningrule})), which is same as Murphy's approach\cite{Murphy2000Combining}. Obviously, if a BOE is collected from a sensor with high reliability, it will have more effect on the final combination results. On the contrary, if a BOE is collected from a sensor with relatively low reliability, it will matter little in the final combination results.
\section{Numerical example}
In this section, a numerical example is illustrated to show the effectiveness of our method. In a target recognition system, five radar sensors have collected five pieces of BOEs shown as follows:
\[\begin{array}{l}
\left( {{R_1},{m_1}} \right) = \left( {\left[ {\left\{ A \right\},0.6} \right],\left[ {\left\{ B \right\},0.15} \right],\left[ {\left\{ {A,C} \right\},0.25} \right]} \right)\\
\left( {{R_2},{m_2}} \right) = \left( {\left[ {\left\{ A \right\},0.5} \right],\left[ {\left\{ B \right\},0.3} \right],\left[ {\left\{ C \right\},0.2} \right]} \right)\\
\left( {{R_3},{m_3}} \right) = \left( {\left[ {\left\{ B \right\},0.95} \right],\left[ {\left\{ C \right\},0.05} \right]} \right)\\
\left( {{R_4},{m_4}} \right) = \left( {\left[ {\left\{ A \right\},0.55} \right],\left[ {\left\{ B \right\},0.25} \right],\left[ {\left\{ {A,C} \right\},0.2} \right]} \right)\\
\left( {{R_5},{m_5}} \right) = \left( {\left[ {\left\{ A \right\},0.6} \right],\left[ {\left\{ B \right\},0.3} \right],\left[ {\left\{ {B,C} \right\},0.1} \right]} \right)
\end{array}\]
The reliability of these sensor reports is 0.55, 0.6, 0.25, 0.45 and 0.5 respectively, which is obtained based on their confidence coefficient curves. Then fusion results and comparison are shown in Table \ref{comparison}.%(the BPA of ${A,C}$ and ${B,C}$ is handled by using Pignistic probability transformation\cite{Smets1994The}).
\begin{table}[!h]
\centering
\caption{Fusion results and comparison}
\label{comparison}
\begin{tabular}{cccc}
\toprule
    & $m\left( A \right)$      & $m\left( B \right)$      & $m\left( C \right)$      \\
\midrule
Classical rule     & 0      & 0.9057 & 0.0943 \\
Murphy's approach  & 0.7971 & 0.2011 & 0.0018 \\
Our method         & 0.9373 & 0.0609 & 0.0018 \\
\bottomrule
\end{tabular}
\end{table}
Four evidences prefer to recognizing the target as $A$. Hence, data from the third sensor is probable to be interfered and incorrect. As can be seen from Table \ref{comparison}, in this situation, our method works better than Murphy's while the classical combining rule does not work. The target can be effectively recognized with our method.

\section{Conclusion}
In summary, we propose a new method to model the reliability of sensor reports. Unlike previous methods, we focus on the properties of a sensor itself. The confidence coefficient curve of a radar sensor is obtained by solving a a quasi time-independent Schr${\ddot o}$dinger equation. The method is used in combining of evidences. The result shows the efficiency of our method.

\section{Acknowledgement}
The work is partially supported by National Natural Science Foundation of China (Grant No. 61671384), Natural Science Basic Research Plan in Shaanxi Province of China (Program No. 2016JM6018), Aviation Science Foundation (Program No. 20165553036), the Fund of SAST (Program No. SAST2016083)

\bibliographystyle{model1-num-names}
\bibliography{myreference}

\end{document}